\documentclass[%
 aip,
 jmp,%
 amsmath,amssymb,
 reprint,%
]{revtex4-1}

\usepackage{graphicx}
\usepackage{dcolumn}
\usepackage{bm}
\usepackage{caption}
\usepackage{subcaption}

\begin{document}

\preprint{AIP/123-QED}

\title[]{Classical Charged Particle Resonance In Induction Field}

\author{Devesh S. Bhosale}
\affiliation{%
 Department of Physics, Stevens Institute of Technology, New Jersey 07302, United States of America 
}
\email{dbhosale@stevens.edu}

\date{\today}
\begin{abstract}
Starting from First Principles, the space charge manipulation of charged particles in an induction field in free space based on an unique Magnetic field strength and its oscillation Frequency relationship is demonstrated numerically and theoretically.
With the dispersion relation in Ion Resonance depending on its frequency of gyration, an AC driven electromagnet based particle resonance has been proposed circumventing the use of Superconducting Permanent Magnets. Complete resonance achieved under the proposed conditions results in a sustained, fixed-frequency particle trajectory that is independent of its speed or drift.
Such oscillation is visualized in a D-Shaped Resonant assembly. The amplitude and the wavelength calculations for the trajectory are demonstrated and the applicability of this unique magnetic field strength and its sinusoidal varying frequency relation in plasma instability is explored.

\end{abstract}
\maketitle

\section{INTRODUCTION}

Resonance phenomena involving charged particles are fundamental across diverse scientific disciplines spanning Particle Accelerators, Spectrometry and Resonant Antennas. This has prompted a surge in research into the gyromotion of charged particles over the past decade, displayed by strong activity in gyrokinetics\citep{PhysRevLett.117.245101}. Amongst the many space charge manipulation that can be imagined, energizing charged ions/electrons and control of plasma profiles have been of great importance, with the development of ICRH and ECRH devices such as the JET Ion Cyclotron Resonance Frequency (ICRF)\cite{PhysRevLett.84.2397}.
  
Efforts to intensify highly charged ion beams have predominantly centred around strengthening superconducting magnets \citep{PhysRevAccelBeams.20.094801} in recent years. In this study we propose a new route for attaining resonance using induction field to produce time-varying magnetic field, instead of permanent magnets employed for such resonance. The peak magnetic field strength of this AC Electromagnet is comparable to the static magnetic field strength of superconducting magnet, adding a new dimension of viability for the realization of particle resonance. 

This paper demonstrates a novel gyromotion of particles using a time varying induction field with its gyrofrequency independent of the particle’s velocity. When the natural gyrofrequency(represented by $\omega_{0}$ in this Article) of such charged particles gyrating in a magnetic field matches the frequency of an external electromagnetic wave, resonance occurs. Much attention has been paid to the study of the oscillations when resonance condition is met, as RF Power absorption from the electromagnetic wave can be efficiently transferred to heat the plasma\cite{10.1063/1.5023631}. 
The dispersion relation for plasma in traditional systems has direct relationship to its Gyrofrequency($\omega_{ci}$) during resonance $\omega-k_{\|} v_{\|}=n\omega_{ci}$ for proton cyclotron resonance \cite{article} and $k_{\|} v_A=\omega_{ci}$ for Ion Cyclotron Resonance  \citep{10.1063/5.0176373} where $\omega_{ci}$ is the cyclotron frequency. As charged particles gain energy from the resonant interaction, they contribute to the excitation of collective plasma modes, leading to increased turbulence, wave-particle interactions, and ultimately, instabilities within the plasma. The production of highly charged ions is strictly governed by this inherent and induced plasma instabilities \citep{PhysRevAccelBeams.21.093402}.

\section{Background Theory}
\begin{figure}[b]
\begin{center}
\includegraphics[width=0.55\textwidth]{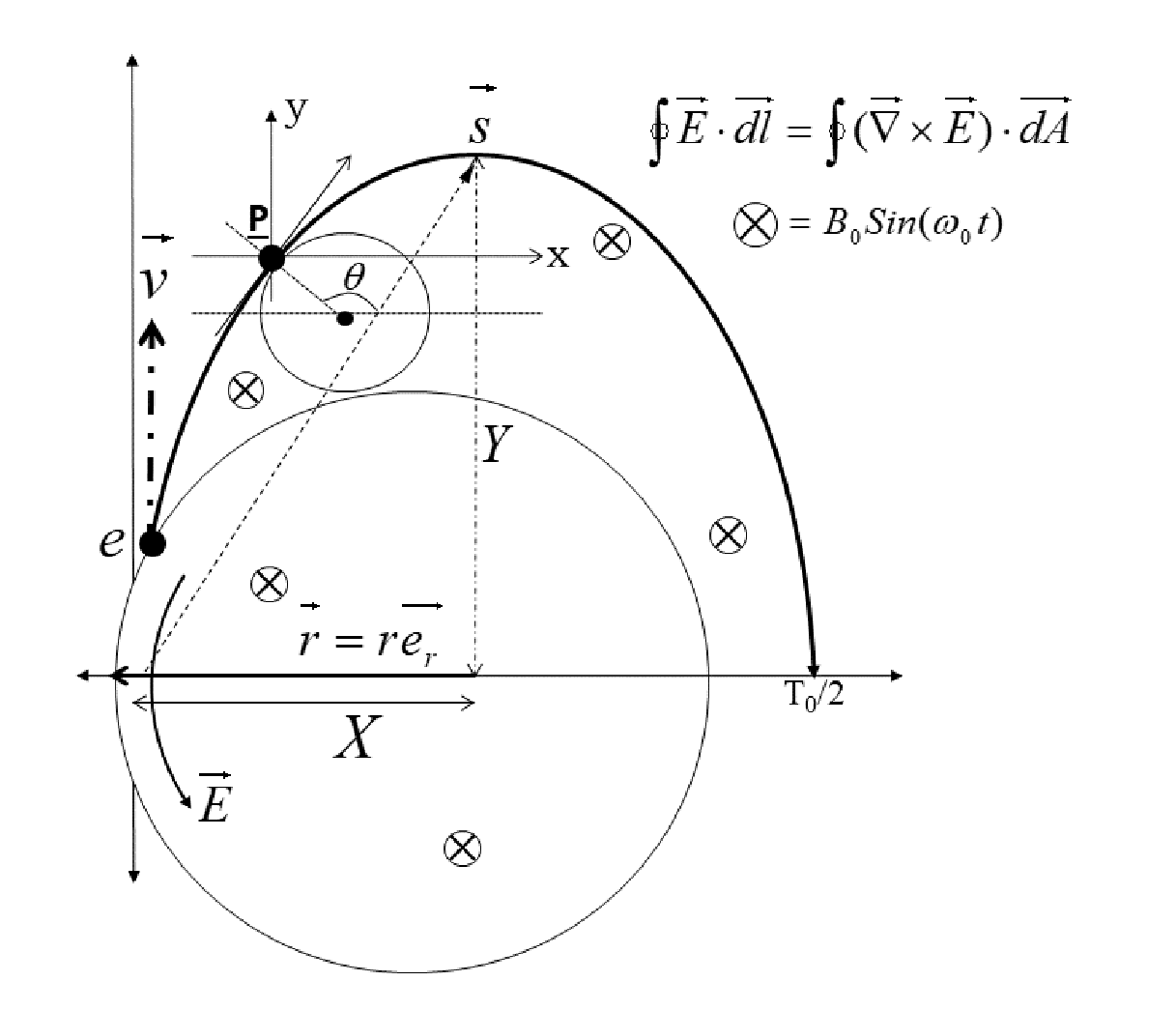}

\end{center}
\caption{\label{fig:fig1} Trajectory of electron with the magnetic field oscillating in and out of the plane.}
\end{figure}
The time-varying sinusoidal magnetic field $B_z(t)=B_0sin(\omega_0t)$ is axially symmetric and applied along the z-axis, while its angular frequency of oscillation being ${\omega_0}$. The induced electric field is calculated using Maxwell's equation and Stoke's Theorem,
\begin{equation}
\oint_{P} \vec{E} \cdot d\vec{l}=-\int_{S} \dot{B}\cdot d\vec{S}
\end{equation}

Giving us the required Electric field as,
\begin{equation}
\vec{E}={E_\theta}=-\frac{1}{2}r\dot{B_z} \hat{e_\theta}
\end{equation}
along the curl defined by surface S normal to z-axis, refer Fig.1. The particle's guiding center may at one point lie on this path $P$ defined along the curl of surface S with radius $r$. Therefore, in this dynamic situation the drift of guiding center, which being along the direction defined by vector $\vec{{E }_\theta}\times \vec{B}$ is equated to time rate change of $r$\citep{seymour1965charged}

\begin{equation}
\dot{r}={v}_d= \frac{\left|\vec{E}\times \vec{B}\right|}{{B}^2}=\frac{{E }_\theta}{B_z}=-\frac{r}{2}\omega_0Cot(\omega_0t)
\end{equation}
Equating (2) and (3) we obtain,
\begin{equation}
\frac{2}{r}\dot{r}+\frac{\dot{B}}{B}=0
\end{equation}
Producing the known condition: \begin{equation}
Br^2=constant=G
\end{equation}

 The velocity of the particle $\vec{v}$ represented in the cylindrical coordinates as
\begin{equation}
\dfrac{d\vec{s}}{dt}=\vec{v}=\dot{r}\hat{e}_r+r\dot{\theta}\hat{e}_\theta+k\hat{z}  
\end{equation}
The particle experiences Lorentz Force for non relativistic case due to the induced electric field E and magnetic field B expressed as:
\begin{equation}
m\dot{\vec{v}}=q(\vec{E}+\vec{v}\times\vec{B}_z) 
\end{equation}

Since the magnetic field is zero at the beginning, the radius corresponding to it is very large. 
The induced electric field follows the curl. On equating the radial components of (7) we obtain 
\begin{equation}
\ddot{r}-r\dot{\theta^2}= \frac{q}{m}{B_z(t)}\cdot r\dot{\theta}
\end{equation}
While the angular component yields,
\begin{equation}
2\dot{r}\dot{\theta}+r\ddot{\theta}= \frac{q}{m}[E_{\theta}-\dot{r}B_z]
\end{equation}
Expanding the angular part by using Equation (2) in Equation (9) and applying Equation (4) to it we immediately arrive at\citep{seymour1965charged}, 

\begin{equation}
\frac{d(r^2\dot{\theta})}{dt}= -\frac{q}{2m}[\frac{d(B_z\cdot r^2)}{dt}]
\end{equation}
Which upon integration produces\citep{seymour1965charged}: 
\begin{equation}
\dot{\theta}= -\frac{q}{2m}{B_z(t)} + C/r^2
\end{equation}
Where $C$ is a real constant. Or,
\begin{equation}
\dot{\theta}= -\frac{q}{2m}{B_z(t)} + D*B_0sin(\omega_0t)
\end{equation}
Where $D$ is the resulting constant by using Equation (5) in (11). Therefore, the integrated form of the above equation gives us the angle swept by particle in the said time about the guiding centre, thereby laying the groundwork for further derivations:
\begin{equation}
\int_{\theta_0}^{\theta}\dot{\theta}= -\int_{0}^{t}\frac{q}{2m}{B_0sin(\omega_0t)}dt + D\int_{0}^{t}B_0sin(\omega_0t)dt
\end{equation}
Where ${\theta_0}$ is the value of ${\theta}$ at $t=0$. Upon substituting $\dot{\theta}$ from Equation (12) in (8) we arrive at, 
\begin{equation}
\ddot{r}+r([\frac{q}{2m}{B_z(t)}]^2 -[D\cdot B_z(t)]^2)=0
\end{equation}
Integrating (13) we obtain,
\begin{widetext}
\begin{equation}
\label{equ}
{\theta}-{\theta_0}=\triangle {\theta}= \frac{qB_0}{2m\omega_0}{cos(\omega_0t)} -\frac{DB_0}{\omega_0}cos(\omega_0t)-\frac{qB_0}{2m\omega_0}+\frac{DB_0}{\omega_0}+A
 \end{equation}
\end{widetext}
Where A is the integration constant. Taking the condition such that when the magnetic field completes its one half cycle, i.e. it completes its positive half cycle in $T_0 /2$, where $T_0$ is the time period of angular frequency $\omega_0$, correspondingly the particle should have traversed from ${\theta}=\pi$-radians to ${\theta}=0$ as can be seen from Fig.2. Substituting $T_0/2$ for $t$ and $-\pi$ for $\triangle {\theta}$ (Since ${\theta_0}=\pi$), in Equation (15), we arrive at:
\begin{equation}
-\pi= -\frac{qB_0}{m\omega_0}+\frac{2DB_0}{\omega_0}+A
\end{equation}
Similarly, in $T_0 /4$ time period the particle should have traversed $-\pi /2$ radians. Hence, we obtain
\begin{equation}
-\pi/2= -\frac{qB_0}{2m\omega_0}+\frac{DB_0}{\omega_0}+A
\end{equation} 
Looking at the above two equations we can immediately see the value of $A=0$. Substituting equation (17) in (15) we arrive at the  \textbf{equation of angular displacement} of particle about its guiding centre for such conditions as:
 \begin{equation}
\triangle {\theta}={\theta}-\pi=- \frac{\pi}{2}[1-{cos(\omega_0t)}]
\end{equation}
\begin{equation}
\large
\underline
{{\theta}=\frac{\pi}{2}[1+{cos(\omega_0t)}]}
\end{equation}
Using (17) we arrive at,
\begin{equation}
D=\frac{q}{2m}+\frac{\pi\omega_0}{2B_0}
\end{equation}
Making appropriate substitution in Equation 14 we form:
\begin{equation}
\ddot{r}=rB_z^2(\frac{q}{m}+\frac{\pi\omega_0}{2B_0})(\frac{\pi\omega_0}{2B_0})
\end{equation}
Differentiating Equation (3) we get an equation for the locus of $r$. Since the above equation holds for all values of $r$, substituting $t=T_0/4$ we get:
\begin{equation}
\ddot{r}=r\frac{\omega_0^2}{2}=rB_0^2(\frac{q}{m}+\frac{\pi\omega_0}{2B_0})(\frac{\pi\omega_0}{2B_0})  
\end{equation}

 \textbf{Thereby forming the necessary relationship} between the peak magnetic field strength $B_0$ and the angular frequency $\omega_0$ at which the particle is undulating:
\begin{equation}
\large
\underline
{
\frac{\pi^2-2}{2\pi}=\frac{qB_0}{m\omega_0}
}
\end{equation}

As an example, the magnetic field strength for the frequency of 250 Mhz comes out to be approx. $0.01118$ Tesla taking electron as our particle. Therefore, our magnetic field function in this case turns out to be: 

$B=0.01118\ast sin(2\pi \ast 250\ast 10^{6}t)$

Compared to the cyclotron frequency of 250 Mhz, the magnetic field required for cyclotron resonance is $0.008931 T$. The equation above indicates that, for a given frequency($\omega_0$), the required magnetic field strength is roughly 1.25 times that of traditional Cyclotron Resonance($\frac{\pi^{2}-2}{2 \pi} \approx 1.25$). This corresponds to the peak magnetic field strength in the oscillating electromagnet($0.01118 T$), in contrast to the static field of a permanent magnet for cyclotron resonance($0.008931 T$). 

Various methods such as using Tank Circuits to reduce the electromagnet's impedance can be used to make the high frequency oscillation at the given field strength feasible\cite{traficante1989impedance}. 
Just as with Permanent Magnets, the magnetic field in electromagnet also has gradient and the field strength is not uniform at all points in the region  . We can see that there is a linear relationship between the field strength and frequency at which particle performs resonance with the electromagnet oscillation frequency $\omega_0$.

\begin{figure*}
\centering
\begin{subfigure}{.54\textwidth}
  \centering
  \includegraphics[width=.9\linewidth]{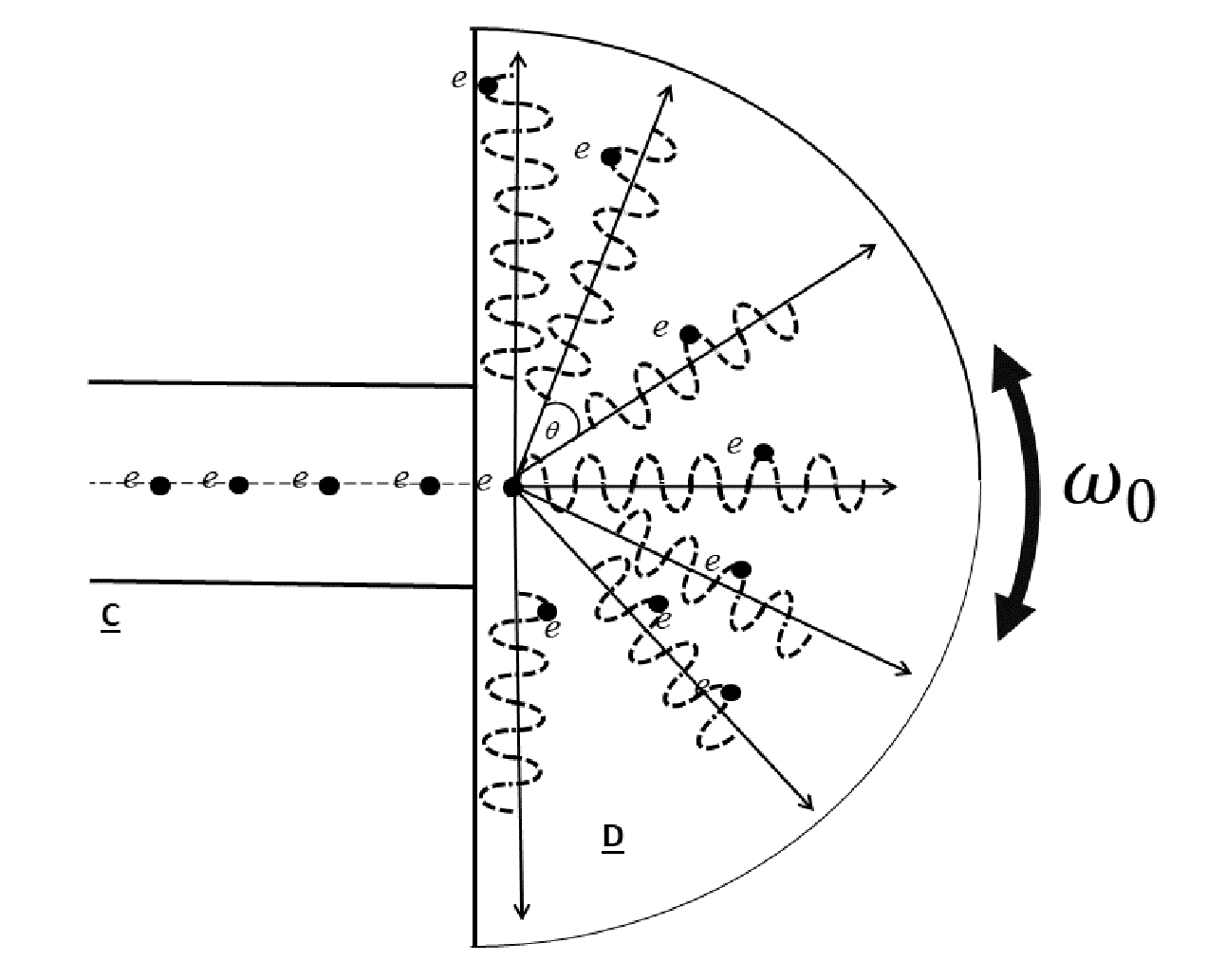}
  \caption{Particle Trajectories}   
  \label{fig:sub1}
\end{subfigure}%
\begin{subfigure}{.54\textwidth}
  \centering
  \includegraphics[width=.98\linewidth]{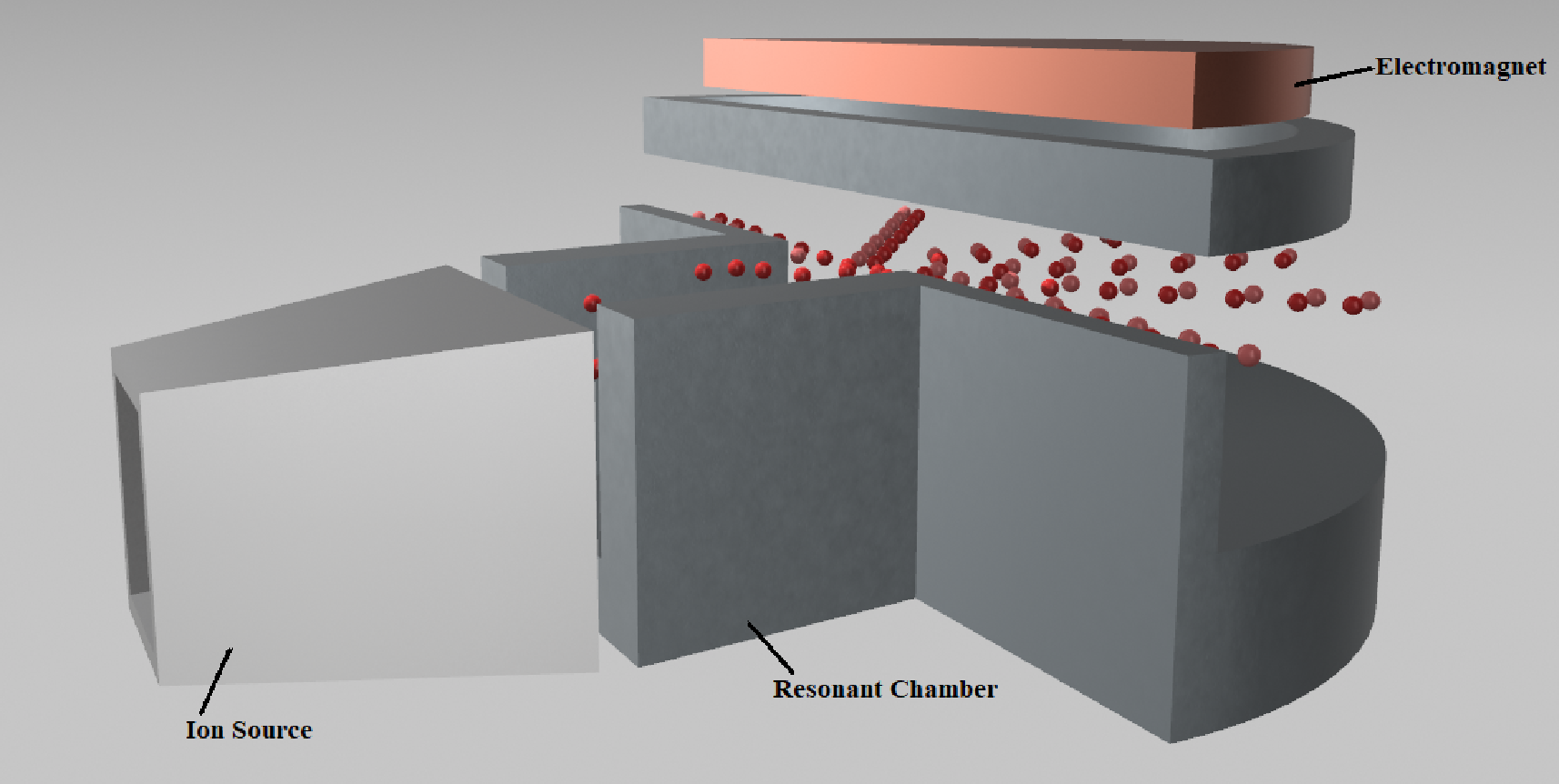}
  \caption{3-d Model of the Assembly}
  \label{fig:sub2}
\end{subfigure}
\begin{subfigure}{.54\textwidth}
  \centering
  \includegraphics[width=.98\linewidth]{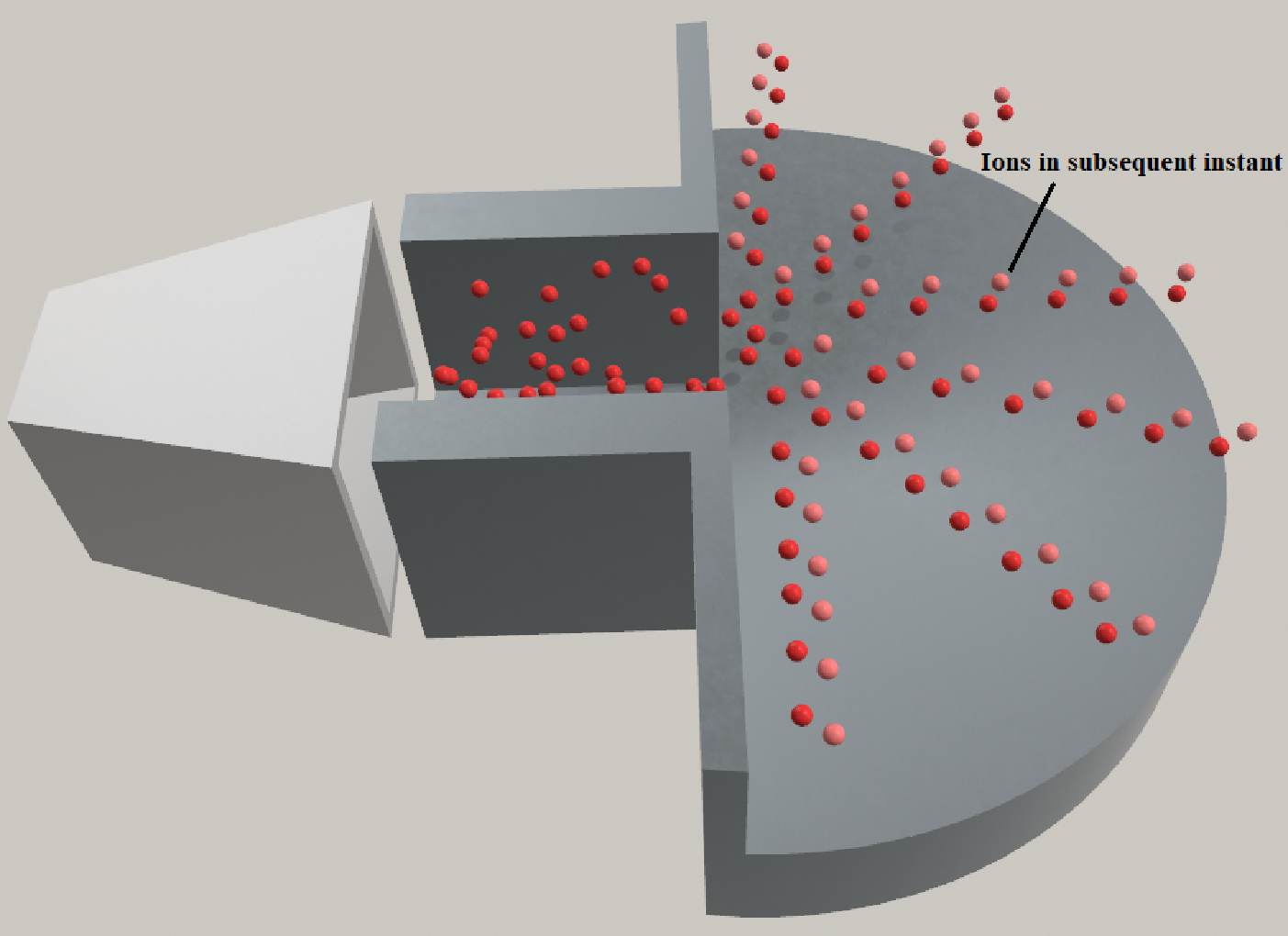}
  \caption{Sectional view of oscillating particles}
  \label{fig:sub3}
\end{subfigure}

\caption{Depiction of the continuous beam of electrons entering the D-section containing the oscillating magnetic field(a) The successive trajectories of electrons with all the particles synchronously oscillating angularly as shown (b) A 3-d model depicting the D-Shaped assembly containing the said electromagnet, source and resonant chamber (c) The oscillating particles' sectional view is depicted, where the light-coloured particles project the subsequent position of the particles.}
\label{fig:test}
\end{figure*}

\section{AMPLITUDE and wavelength CALCULATION}
Equation (6) can be resolved from cylindrical to Cartesian coordinate system:
\begin{equation}
\vec{v}=\dot{r}[cos(\theta)\hat{i}+sin(\theta)\hat{j}]+r\dot{\theta}[-sin(\theta)\hat{i}+cos(\theta)\hat{j}]+\vec{v_z}
\end{equation}
We take $v_z=0$ for the simplicity of calculations. From equation (3) and (19) to the above equation we form:

\begin{widetext}
\begin{equation}
\label{equ}
\large
\vec{v}={r}\frac{\omega_0}{2}([-Cot(\omega_0t)cos(\theta)+\pi sin(\omega_0t)sin(\theta)]\hat{i}-[Cot(\omega_0t)sin(\theta)+\pi sin(\omega_0t) cos(\theta)]\hat{j})
 \end{equation}
\end{widetext}
It is to be noted that that the polar coordinates are defined with respect to the guiding center formed when the particle enters the oscillating magnetic field. Using the constant from Equation (5) we write Equation (25) as,
\begin{widetext}
\begin{equation}
\label{equ}
\dfrac{d\vec{s}}{dt}=\vec{v}=\frac{\sqrt{G}\cdot \omega_0}{2\sqrt{B_0}}([-\frac{cos(\omega_0t)}{sin(\omega_0t)^{3/2}}cos(\theta)+\pi \sqrt{sin(\omega_0t)}sin(\theta)]\hat{i}-[\frac{cos(\omega_0t)}{sin(\omega_0t)^{3/2}}sin(\theta)+\pi \sqrt{sin(\omega_0t)} cos(\theta)]\hat{j})
\end{equation}
\end{widetext}
As magnetic field does no work on charged particle, the induced Electric field  will be the one doing work on it\cite{PhysRevE.78.032102}. In order to find the value of constant $G$ we use the work-energy theorem while taking electron as our particle,
\begin{equation}
\Delta K.E.= e \int \vec{E}\cdot\vec{ds} 
\end{equation} 
Using Equation (2), (5) and (6) we can form the integral as, 
\begin{equation}
e\int_{0}^{T_0/4}\vec{E}\cdot\vec{ds} =e\int_{0}^{T_0/4} \frac{1}{2}r^2\dot{B_z}\dot{\theta}  dt =\frac{\pi}{4}eG\omega_0 
\end{equation}
While the velocity from Equation (26) at $t=T_0/4$  is,
\begin{equation}
v_0=\frac{\sqrt{G}\cdot \omega_0}{2\sqrt{B_0}} \pi\hat{i} 
\end{equation}
Applying work-energy theorem for a non-relativistic speed of particle we form,
\begin{equation}
\frac{\pi}{4}eG\omega_0=\frac{1}{2}mv_0^2-\frac{1}{2}mu^2  
\end{equation}
where $u$ is the initial velocity with which electron has been projected with and $v_0$ the velocity at $T_0/4$, finally giving us the value of G as.
\begin{equation}
G=u^2/(\frac{\omega_0^2\pi^2}{4B_0}-\frac{e \pi\omega_0}{2m}) 
\end{equation}
 Substituting this $G$ in Equation (26) with the displacement vector $\vec{ds}$ summing up from origin and integrating it over to $T_0/4$ we obtain,
\begin{equation}
\vec{S}=\frac{\sqrt{G}\cdot \omega_0}{2\sqrt{B_0}}(43.0942\hat{i}+1.5552\hat{j})
\end{equation}
 wherein the required amplitude is $\frac{\sqrt{G}\cdot \omega_0}{\sqrt{B_0}}(0.7776)$ and while the required wavelength is $4\sqrt{X^2+Z^2}$, with $Z$ forming an equivalent to the pitch length in Ion Cyclotron motion, here with $v_z=0$ its simply $\frac{\sqrt{G}\cdot \omega_0}{\sqrt{B_0}}(86.1884)$.

\section{Lagrangian and Continuous Beam}

The resulting single particle trajectory is visually similar to that of electrons in Wiggler. When a continuous beam of electrons enters the D-section containing the said magnetic field $B_z$, the successive particle trajectories has radial distribution as can be seen from Fig.2. This occurs because particle trajectory begins from the corresponding Larmor Orbit of the magnetic field strength at the instant it enters the D-section.
The angular displacement $\theta$ from the Equation 19 is also the same angular displacement between successive electron travel paths. Its to be noted that there is a synchronous oscillation of all the particles/plasma present in the D-section along its angular direction as indicated in Fig.2. The frequency of this resonance is $\omega_0$, which draws parallels to the cyclotron frequency $\omega_{ci}$ in permanent magnet based resonance.

The Lagrangian in polar coordinates for such particle dynamics is:
\begin{widetext}
\begin{equation}
\label{equ}
\large
L = \begin{aligned} & \frac{mr^{2}} {2} \left[\frac{\omega_{0}^{2}}{4} \cot ^{2}(\omega_{0}t)+\frac{\pi^{2}}{4} \omega_{0}^{2} \sin ^{2}\left(\omega_{0} t\right)\right] -e B_{0} \omega_{0} \cos \left(\omega_{0} t\right) \pi r^{2}-\frac{B_{0}^{2} \sin ^{2}\left(\omega_{0} t\right)* \pi r^{2} h}{2\mu}\end{aligned}
\end{equation}
\end{widetext}
where h represents the height of the volumetric distribution of magnetic field in the assembly, while ${\mu}$ is the effective permeability of the medium. A reproduction of the external magnetic field in the Vlasov equation along with the stated variables and with an appropriate distribution function can help identify the resonance conditions, forming the basis for applications in plasma instability. Coupling of RF power into the Plasma using RF Antennas, although not displayed in the assembly above and an appropriate accountability for the relativistic velocities is left to the experts for future discussions.

\section{Conclusion}

 In this paper, we have demonstrated a novel resonance condition for charged particles. Particle dynamics generated by such process has been visualized in the schematics. By extending the model to the case of non-
relativistic velocities, we have obtained a simple procedure to achieve a source of RF Power, increase the ion kinetic energy and consequently, increase the plasma temperature for ICRH or ECRH.  
 
 The invariance of Kinetic Energy for the non-relativistic velocities has been displayed in the absence of radiation losses. While in Magnetic pumping, heating or cooling, capabilities are limited as the field cannot be ramped up or down indefinitely, it is ideal if particles can be heated or cooled in a periodically-varying magnetic field\citep{osti_1000551}. The next-order kinetic energy variation by increasing pumping frequency can be calculated henceforth from this theory. The control of orbit is still limited by maximal magnetic field in the pursuit of higher energy ions\citep{osti_1000551}, the current method attempts to overcome this by offering higher magnetic field strength employing electromagnets for resonance. Principally noting the objectives of Synchrotron in varying the magnetic field and its frequency to generate high energy particles\cite{PhysRev.68.143}, the present theory addresses the same with a reduction in cost by omitting the use of superconducting magnets. Its applications can be explored in magnetic confinement and sub-harmonic heating and cooling. This approach and the resulting phenomenology may prove valuable in plasma and accelerator physics.

\begin{acknowledgments}

I would like to thank Late Prof. Predhiman Kaw from Institute of Plasma Research, Dr. R S Shinde and Mr. Subrata Das RRCAT, Indore for valuable comments.

\end{acknowledgments}

\nocite{*}

\bibliography{apssamp}

\end{document}